\documentclass[12pt]{article}
\usepackage{a4wide}
\usepackage{amssymb}
\usepackage{graphicx}
\begin{document}
{\renewcommand{\thefootnote}{\fnsymbol{footnote}}
  \begin{center}
    {\LARGE  Tunneling dynamics of an oscillating universe model}\\
    \vspace{1.5em} 

    Martin Bojowald\footnote{e-mail address: {\tt
        bojowald@psu.edu}}
    and Pip Petersen\footnote{e-mail address:
      {\tt sxp1171@case.edu}}\footnote{New address: Department of Physics, Case Western
      Reserve University, 2076 Adelbert Road, Cleveland, OH 44106, USA}
    \\
    \vspace{0.5em}
    Department of Physics,\\
    The Pennsylvania State
    University,\\
    104 Davey Lab, University Park, PA 16802, USA\\
    \vspace{1.5em}
\end{center}
}

\setcounter{footnote}{0}

\begin{abstract}
  Quasiclassical methods for non-adiabatic quantum dynamics can reveal new
  features of quantum effects, such as tunneling evolution, that are harder to
  reveal in standard treatments based on wave functions of stationary
  states. Here, these methods are applied to an oscillating universe model
  introduced recently. Our quasiclassical treatment correctly describes
  several expected features of tunneling states, in particular just before and
  after tunneling into a trapped region where a model universe may oscillate
  through many cycles of collapse and expansion. As a new result, the
  oscillating dynamics is found to be much less regular than in the classical
  description, revealing a succession of cycles with varying maximal volume
  even when the matter ingredients and their parameters do not change.
\end{abstract}

\section{Introduction}

Tunneling effects are relevant in oscillating universe models obtained by
recasting Friedmann dynamics in terms of the motion of the scale factor in a
potential \cite{OscillatingFriedmann,OscillatingSimple}. Classical
oscillations may then become unstable in quantum cosmology if at least one of
the relevant potential barriers around the oscillation region are of finite
height and width. Such instabilities have been studied in
\cite{OscillatingInstability,OscillatingEternal} and, with an emphasis on tunneling,
in \cite{OscillatingTunnel,OscillatingCollapse,OscillatingRate}.

Here, we demonstrate that not only the traditional tunneling probability
familiar from stationary problems in standard quantum mechanics is of interest
and computable, but also a more detailed picture of time-dependent tunneling
dynamics. The methods we use, given by canonical effective descriptions of
evolving quantum states based on the non-adiabatic dynamics of moments, have
already proven useful in other fields, for instance by shedding light on the
question of tunneling or traversal times \cite{Bosonize,EffPotRealize,Ionization} in atomic
physics.

The quasiclassical method we apply here reformulate quantum dynamics of states
as a coupled system of ordinary differential equations for expectation values
of a basic set of operators together with higher moments. Such extended
systems of equations could also be obtained classically if a distribution of
like objects is considered instead of a single point particle
\cite{ClassMoments,MomentsQuartic}. Quantum dynamics, however, not only
introduces additional corrections in these equations for non-zero $\hbar$, it
also gives the statistical degrees of freedom described by moments of a state
a more fundamental role because they are then unavoidable. In general, the
quantum state space is infinite-dimensional and hard to parameterize
completely, but we will see that a single additional quantum parameter,
identified with the size of quantum fluctuations and also used to parameterize
higher moments in a suitable way, is sufficient to reveal interesting new
features, in particular of the tunneling dynamics. 

Even this restriced extension to one additional quantum degree of freedom and
its canonically conjugate momentum reveals a quantum dynamics that is much
more complicated than the regular classical one, and possibly chaotic. The
classical dynamics of an isotropic universe as formulated in
\cite{OscillatingFriedmann,OscillatingSimple} makes use of a 1-dimensional
potential and is therefore guaranteed to be integrable.

Our extension by a single quantum parameter suffices to complicate the
dynamics and possibly introduce chaotic features. We only provide
circumstantial evidence for chaos in this paper and focus on a qualitative
description of generic features of the extended dynamics. In particular, the
additional parameter, compared with the classical formulation, implies that
the universe, generically, enters different cycles of expansion and collapse
with different initial values of the quantum parameter. Since this parameter
couples to the evolution of the scale factor, the latter reaches different
maximum values in different cycles. Properties of cycles may therefore vary
even if the matter ingredients and their parameters remain the same. In this
way, a single model can give rise to a larger variety of universe cycles and
more easily accomodate properties of a single observed universe.

\section{Oscillating model}

We start with the specific potential derived from an oscillating universe model
introduced in \cite{OscillatingSimple}. The model is spatially isotropic, has
positive spatial curvature, and an energy density given by
\begin{equation}
  \rho(a)=\Lambda+\frac{\sigma}{a}+ \rho_{\phi}
\end{equation}
including a negative cosmological constant, $\Lambda<0$, a matter density
contribution $\sigma/a$ with a positive constant $\sigma>0$, as well as the
energy density $\rho_{\phi}$ of a scalar field $\phi$. The Friedmann equation
therefore reads
\begin{equation}
  \frac{\dot{a}^2}{a^2} + \frac{k}{a^2}= \frac{8\pi
    G}{3}\left(\Lambda+\frac{\sigma}{a}+\rho_{\phi}\right)\,.
\end{equation}

As described in \cite{OscillatingRate} for this model, the inclusion of a
free, massless scalar field $\phi$ is useful because it implies two degrees of
freedom, $a$ and $\phi$, that can evolve with respect to each other. Thus
avoiding any reference to a time coordinate, which would not be subject to
quantization, the scalar degree of freedom will help with the interpretation
of dynamics in quantum cosmology following \cite{Blyth}. Note, however,
that this relational evolution by itself does not solve the problem of time in
quantum cosmology \cite{KucharTime,IshamTime,AndersonTime} because it requires
a specific choice of time degree of freedom, $\phi$, and is not guaranteed to
provide quantum results independent of the choice of time
\cite{TwoTimes,MultChoice,ClocksDyn,BianchiInternal,Switching,SingClock,ClockDep}.

The scalar being free of self-interactions and massless, its energy
contribution is
\begin{equation}
  \rho_{\phi}= \frac{p_{\phi}^2}{2a^6}
\end{equation}
with the momentum $p_{\phi}$ of $\phi$.
Since there is no explicit $\phi$-dependence, the scalar evolution equations
imply that $p_{\phi}$ is a conserved quantity and that $\phi$ is monotonic
with respect to any time coordinate as long as $p_{\phi}\not=0$. Therefore,
$\phi$ itself may be used as a global time coordinate to formulate quantum
evolution of wave functions. Our methods will, however, be quasiclassical, as
described in more detail below, and do not require the choice of a matter
degree of freedom as time. Nevertheless, we keep the scalar energy density
because it affects the dynamics of the scale factor through its appearance in
the Friedmann equation. It may be considered a simple version of matter
contributions not included in $\Lambda$ and $\sigma/a$. Our specific results
will only depend on the general feature that such energy contributions should
be positive.

The curvature parameter $k$ is positive by assumption and would equal $k=1$ in
the standard normalization of $a$ if one assumes that all of isotropic space
at any given time can be described as a complete 3-sphere. More generally, one
may assume $0<k<1$ if the isotropic dynamics is interpreted as describing a
collection of independent isotropic patches, approximating an
inhomogeneous universe. (As per \cite{OscillatingEternal}, there are certain
string effects that could also reduce the value of an effective $k$ to be
below one.) According to the Belinskii--Khalatnikov--Lifshitz (BKL) scenario
\cite{BKL}, the generic cosmological dynamics close to a spacelike singularity
may indeed be approximated by a collection of independent homogeneous patches,
although the generic dynamics would suggest a certain anisotropic geometry for
each patch. As usual, the isotropic Friedmann equation serves as a simple
first approximation to anisotropic but still homogeneous collapse or
expansion. The value of $k$ then determines the coordinate size of each patch
as a fraction $k^{3/2}$ of the unit 3-sphere volume. Details of the BKL
scenario show that, classically, homogeneous spatial patches close to a
spacelike singularity are asymptotically small without a non-zero lower
bound. The near-big bang behavior should therefore be described by small
$k$. Since small $k$ correspond to microscopic patches, their dynamics is
usually more sensitive to various quantum effects than the dynamics of a
single macroscopic space with $k=1$ \cite{Infrared}.

The patch model is particularly relevant for tunneling questions because it
provides meaning to a tunneling probability, or to our description below in
terms of expectation values and moments of a state. These statistical concepts
require an ensemble of universe models, which in the patch picture can be
individual constituents of the single universe that we are able to observe.

\subsection{Potential}

Given the sign choice of the cosmological constant, the Friedmann equation can
be rewritten as the zero-energy condition
\begin{equation} \label{FriedmannEnergy}
 0= \dot{a}^2+\omega^2(a-\gamma/\omega)^2+k-\gamma^2-\frac{\tilde{p}^2}{a^4} =
 \dot{a}^2+ U_{\rm harmonic}(a)- \frac{\tilde{p}^2}{a^4}
\end{equation}
where
\begin{equation} \label{Uharmonic}
  U_{\rm harmonic}(a)= \omega^2(a-\gamma/\omega)^2+k-\gamma^2
\end{equation}
is, up to constant shifts, a standard harmonic-oscillator potential with
\begin{equation} \label{omegagamma}
  \omega=\sqrt{-\frac{8\pi G\Lambda}{3}} \quad\mbox{and}\quad
  \gamma=\sqrt{-\frac{2\pi G\sigma^2}{3\Lambda}}\,.
\end{equation}
The scalar density provides an anharmonic contribution determined by the constant
\begin{equation}
  \tilde{p}=\sqrt{\frac{4\pi G}{3}} p_{\phi}\,.
\end{equation}

The canonical formulation of the model does not look quite the same as the
standard harmonic oscillator because the canonical momentum of $a$, according
to general relativity, is not simply a constant times $\dot{a}$ but rather
given by
\begin{equation}
 p_a=-\frac{3}{4\pi G} a \dot{a}\,.
\end{equation}
(Heuristically, as explained in more detail in \cite{Foundations}, since the
universe has no matter-independent mass that could be used to form a momentum
from $\dot{a}$, an additional factor of $a$ in combination with Newton's
constant $G$ is required.) Upon replacing $\dot{a}$ in (\ref{FriedmannEnergy})
with $p_a$, the $\omega$-term in the canonical potential of
\begin{equation} 
 0= \frac{16}{9}\pi^2G^2 p_a^2+ a^2 U_{\rm harmonic}(a)- \frac{\tilde{p}^2}{a^2}
\end{equation}
is therefore quartic in $a$.

The scale factor in a strict sense takes values in a semi-bounded range, given
by positive numbers. Its canonical quantization therefore requires a suitable
treatment of a phase space with a boundary, as undertaken for instance in
\cite{AffineSmooth,AffineSing,SpectralAffine,MixAffine} by applying methods
from affine quantum gravity \cite{AffineQG,AffineQG2}. Alternatively, one may
first perform a canonical transformation from $(a,p_a)$ to a canonical pair,
$(\alpha,p_{\alpha})$, suitable for a logarithmic scale factor $\alpha$. As in
\cite{OscillatingRate}, we use the logarithmic scale factor
\begin{equation} \label{alpha}
  \alpha=\ln(\omega\gamma a)
\end{equation}
making use of the two parameters (\ref{omegagamma}) that characterize the
harmonic potential (\ref{Uharmonic}). The definition (\ref{alpha}) is
turned into a canonical transformation if it is accompanied by
\begin{equation}
  p_{\alpha}=ap_a= -\frac{3}{4\pi G} a^2\dot{a}\,.
\end{equation}
The canonical energy equation for $(\alpha,p_{\alpha})$ therefore reads
\begin{equation} \label{alphaConstraint}
  0=\frac{16}{9} \pi^2G^2 p_{\alpha}^2+ \frac{1}{\omega^4\gamma^4}
  e^{4\alpha} U_{\rm harmonic}(a(\alpha))- \tilde{p}^2\,.
\end{equation}
Defining
\begin{equation}
  \beta=\frac{4\pi G}{3} \omega^2\gamma^2 \quad\mbox{and}\quad
  p=\frac{3}{4\pi G} \tilde{p}\,,
\end{equation}
we finally obtain the basic dynamical equation
\begin{equation} \label{constraint}
  0=p_{\alpha}^2+U_p(\alpha)
\end{equation}
with the potential
\begin{equation} \label{Up}
  U_p(\alpha) = \frac{e^{4\alpha}}{\beta^2} 
  \left(k-2e^{\alpha}+\frac{e^{2\alpha}}{\gamma^2} \right) -p^2\,.
\end{equation}

\begin{figure}
\begin{center}
  \includegraphics[width=13cm]{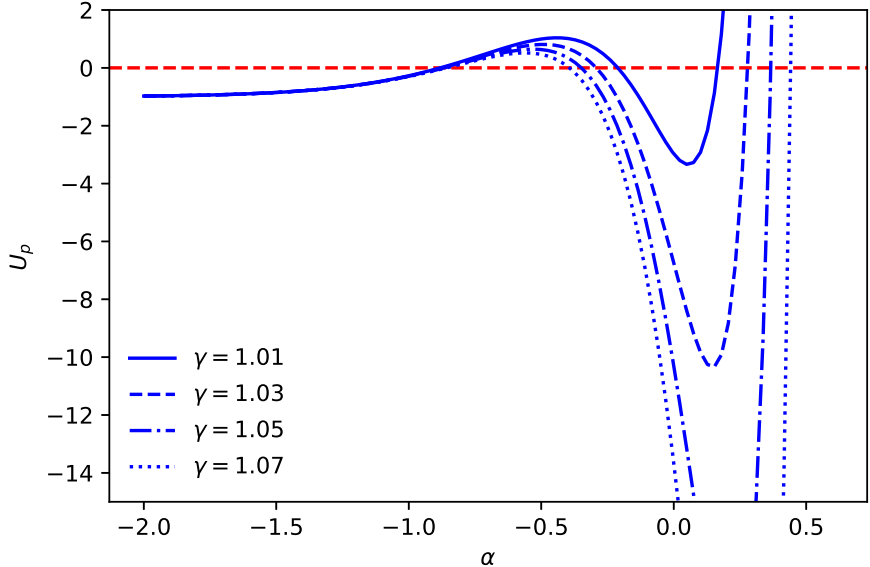}
  \caption{The potential (\ref{Up}) for different values of $\gamma$:
    $\gamma=1.01$ (solid), $\gamma=1.03$ (dashed), $\gamma=1.05$ (dash-dotted)
    and $\gamma=1.07$ (dotted). The dependence on $\gamma$ is rather sensitive
    and determines the width and depth of confined regions with classical
    oscillations. The other parameters used in this plot are $k=1$, $\beta=1$
    and $p=1$. \label{f:Up}}
\end{center}
\end{figure}

\begin{figure}
\begin{center}
  \includegraphics[width=13cm]{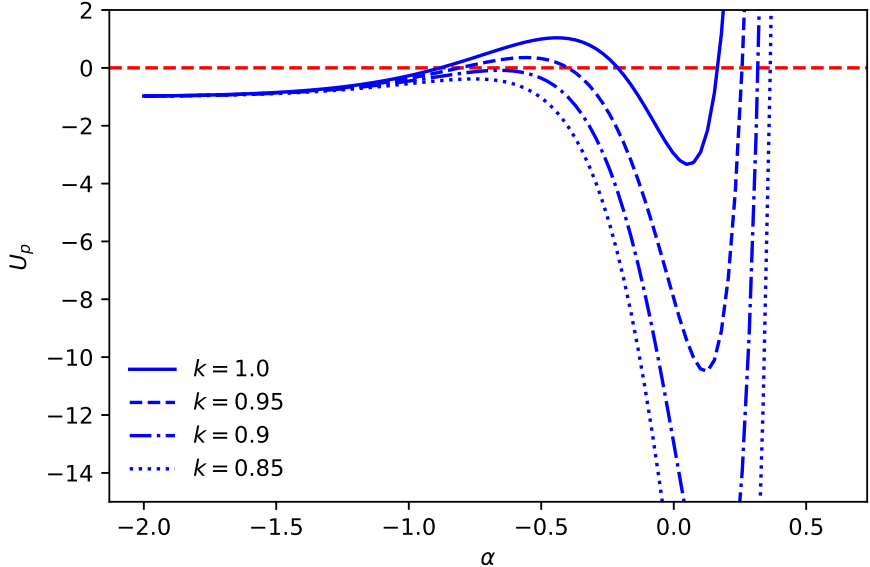}
  \caption{The potential (\ref{Up}) for different values of $k$: 
    $k=1.0$ (solid), $k=0.95$ (dashed), $k=0.9$
    (dash-dotted) and $k=0.85$ (dotted). The other parameters are $\gamma=1.01$, $\beta=1$ and
    $p=1$. \label{f:Upk}}
\end{center}
\end{figure}

This potential is illustrated in Figs.~\ref{f:Up} and \ref{f:Upk} for
different values of $\gamma$ and $k$, respectively. (The influence of $\beta$
on the potential is easy to see because this parameter simply appears in a
multiplier of the potential, except for the constant shift by $-p^2$.)  The
dependence on both parameters is rather sensitive and determines the width and
depth of the regions of classical oscillations. Moreover, choosing smaller $k$
at fixed $\gamma$ reduces the height of the barrier and can, for non-zero $p$,
reduce the maximum to a value below zero, making it possible for the classical
universe to collapse into a singularity ($\alpha\to-\infty$). The BKL-type
fragmentation of space modeled by homogeneous patches close to a spacelike
singularity, which requires smaller and smaller $k$ as the universe collapses
in order to maintain the homogeneous approximation, is therefore a new source
of instability of oscillating universe models. Since we are mainly interested
in analyzing the dynamics of quantum tunneling, assuming that it is relevant
for the instability because the classical model would be stable, we will work
with the value $k=1$ in what follows.

Before we introduce quantum effects, we mention that evolution in proper time
is generated by the constraint via Hamilton's equations. The relevant Hamilton
function, as usual, is an energy expression, which is not the same as the
standard kinetic energy plus an effective potential (\ref{Up}) used to
visualize the motion in terms of barriers and allowed regions. The proper-time
Hamiltonian differs from the right-hand side of (\ref{alphaConstraint}) by
multiplication with a suitable power of $a$ or $\exp(\alpha)$, up to
constants, because we have been multiplying the matter energy with several
such factors in the process of performing transformations. Tracing back all
these steps, proper-time evolution should be generated by the right-hand side
of (\ref{alphaConstraint}) times $\exp(-3\alpha)$. Up to constant factors,
this multiple turns the $p$-term into the energy of a free, massless scalar
field and therefore provides the correct generator of evolution. We may still
use the potential landscape according to (\ref{Up}) to visualize the dynamics,
but for quantitative estimates of time durations we should keep in mind that
proper-time evolution is slowed down for larger $\alpha$ compared with what
the potential would suggest. (We noticed that including the exponential factor
of $\exp(-3\alpha)$ for proper-time dynamics complicates the numerical
solution of differential equations because the factor changes quickly in some
regions of the relevant phase space.)

\subsection{Canonical effective methods}

For a semiclassical description of tunneling dynamics, it is important to use
non-adiabatic methods that allow one to go beyond stationary states. A
suitable canonical formulation can be obtained by writing wave-function
dynamics in terms of a dynamical system for expectation values of basic
operators, such as $\hat{\alpha}$ and $\hat{p}_{\alpha}$, in a state coupled
to fluctuations and higher moments, generically
\begin{equation}
  \Delta(\alpha^ap_{\alpha}^b)=\langle(\hat{\alpha}-\langle\hat{\alpha}\rangle)^a
  (\hat{p}_{\alpha}-\langle\hat{p}_{\alpha}\rangle)^b\rangle_{\rm symm}
\end{equation}
in completely symmetric, or Weyl, ordering. A phase-space structure is
obtained for these variables by defining the Poisson bracket
\begin{equation}
  \{\langle\hat{A}\rangle,\langle\hat{B}\rangle\}=
  \frac{\langle[\hat{A},\hat{B}]\rangle}{i\hbar}
\end{equation}
and extending it to moments by using the Leibniz rule \cite{EffAc,Karpacz}.

While $\{\langle\alpha\rangle,\langle\hat{p}_{\alpha}\rangle\}=1$ according to
this definition, the Poisson bracket of moments is non-canonical. (For
instance, $\{\Delta(\alpha^2),\Delta(p_{\alpha}^2)\}=4\Delta(\alpha
p_{\alpha})$.) The transformation from the 3-dimensional space of second-order
moments to new variables $(s,p_s,U)$, defined by
\begin{eqnarray}
  \Delta(\alpha^2) &=& s^2 \label{s}\\
  \Delta(\alpha p_{\alpha}) &=& sp_s\\
  \Delta(p_{\alpha}^2) &=& p_s^2+\frac{U}{s^2}\,, \label{ps}
\end{eqnarray}
turns out to imply a canonical bracket $\{s,p_s\}=1$ while
$\{U,s\}=0=\{U,p_s\}$. (These canonical variables have been introduced several
times independently for various studies of semiclassical dynamics
\cite{VariationalEffAc,GaussianDyn,EnvQuantumChaos,QHDTunneling,CQC,CQCFieldsHom}.)
The parameter $U$, which equals the uncertainty expression
$\Delta(\alpha^2)\Delta(p_{\alpha})^2-\Delta(\alpha p_{\alpha})^2$ as a
consequence of the mapping (\ref{s})--(\ref{ps}), is therefore a Casimir
variable of the Poisson manifold. That is, it has vanishing Poisson brackets
with basic expectation values and all second-order moments and is conserved by
any canonical dynamics of these variables. Heisenberg's uncertainty relation
implies the lower bound $U\geq \hbar^2/4$.

Given a Hamilton operator $\hat{H}$, a canonical effective Hamiltonian can be
derived by inserting the mapping (\ref{s})--(\ref{ps}) in the expectation
value $\langle\hat{H}\rangle$. For us, the relevant expression is given by the
constraint (\ref{constraint}) with the non-polynomial potential (\ref{Up}). A
Taylor expansion of the potential --- formally in
$\Delta\hat{\alpha}=\hat{\alpha}-\langle\hat{\alpha}\rangle$, after inserting
$\langle\hat{\alpha}\rangle+ \Delta\hat{\alpha}$ in the quantum operator
$\langle U_p(\hat{\alpha})\rangle$ --- implies the moment-corrected constraint
\begin{equation} \label{Taylor}
  0=\langle \hat{p}_{\alpha}\rangle^2+ \Delta(p_{\alpha}^2)+
  U_p(\langle\hat{\alpha}\rangle)+ \sum_{n=2}^{\infty}\frac{1}{n!}
  \frac{{\rm d}^nU_p(\langle\hat{\alpha}\rangle)}{{\rm d}\langle\hat{\alpha}\rangle^n} \Delta(\alpha^n)
\end{equation}
with an infinite series of higher moments. Including only moments of second
order and using (\ref{s})--(\ref{ps}) as well as the simplified notation
$\alpha=\langle\hat{\alpha}\rangle$ and
$p_{\alpha}=\langle\hat{p}_{\alpha}\rangle$, we obtain the canonical
expression
\begin{equation} \label{USecond}
  0=p_{\alpha}^2+p_s^2+ \frac{U}{s^2}+ U_p(\alpha)+ \frac{1}{2} U_p''(\alpha)
  s^2
\end{equation}
for our semiclassical constraint. 

Tunneling processes rely on higher-order moments because wave packets not only
spread out, as described by the variance $\Delta(\alpha^2)$, but also split up
into reflected and tunneled wave packets. An extension of the canonical
mapping (\ref{s})--(\ref{ps}) to higher orders is challenging, not the least
because the dimension of the Poisson manifold quickly increases when new
moments are included as independent degrees of freedom. For explicit mappings
to canonical variables for moments of third and fourth order, see
\cite{Bosonize,EffPotRealize}.

Instead of using a full mapping to higher orders, closure conditions have
proven useful in studies of tunneling. Such conditions present an approximate
description of higher-order moments in terms of lower-order parameters such as
$s$, without including additional degrees of freedom for them. An example
would be a Gaussian closure because for a Gaussian state, all moments are
determined by second-order ones. A slightly different example that is
algebraically simpler in effective potentials is the all-orders closure
proposed in \cite{Ionization}, where
\begin{equation} \label{Closure}
  \Delta(\alpha^n)=s^n
\end{equation}
for even $n$ while $\Delta(\alpha^n)=0$ for odd $n$. With this closure, the
whole series in (\ref{Taylor}) can be summed explicitly to obtain the simple
constraint
\begin{equation} \label{UAll}
   0=p_{\alpha}^2+p_s^2+\frac{U}{s^2}
   +\frac{1}{2}\left(U_p(\alpha+s)+U_p(\alpha-s)\right)\,.
\end{equation}
A similar expression of effective potentials for certain classes of states has
also been derived from Wigner functions \cite{WignerSemiclass}. 

\begin{figure}
\begin{center}
  \includegraphics[width=16cm]{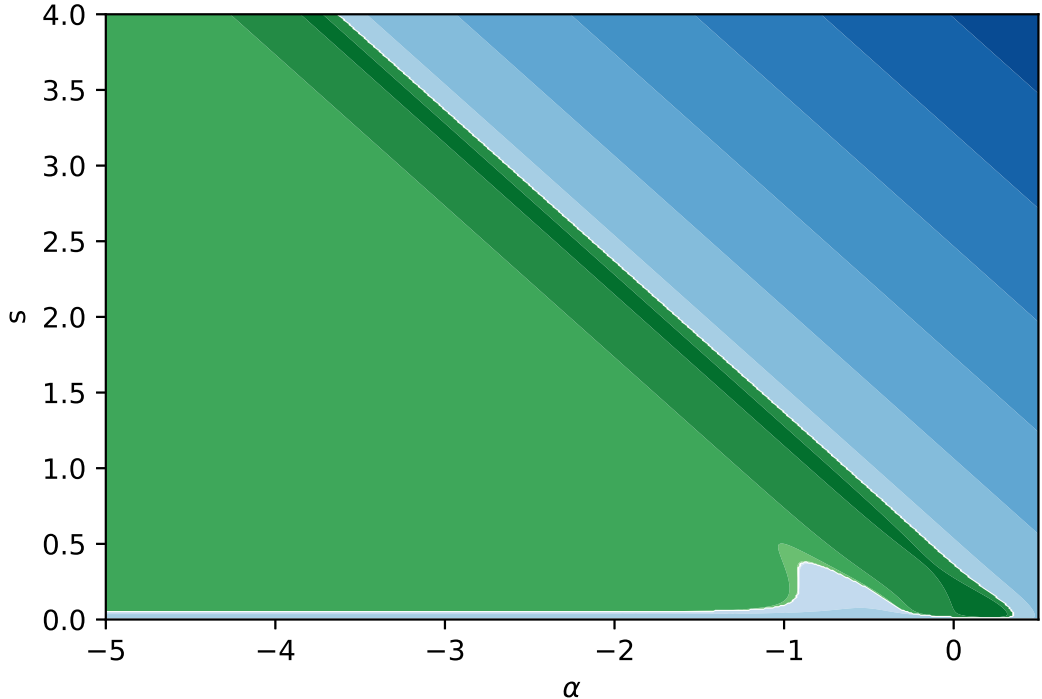}
  \caption{Logarithmic plot of the quantum potential in (\ref{UAll}) for
    $\gamma=1.05$, $\beta=0.1$, $p=1.0$ and $4U=10^{-2}$. The classical
    barrier around $\alpha=-0.5$ is reduced in the $s$-direction while it
    moves to smaller $\alpha$. Around $s=0.5$ at $\alpha\approx -1$, the
    barrier height falls below zero, such that the left-most region connects
    with a channel of negative potential that ends at the classically confined
    region. Tunneling out of the classically confined region can therefore be
    described quasi-classically by motion in an extended phase-space,
    by-passing the classical barrier while maintaining energy
    conservation. The quantum variable $s$ has to grow sufficiently large
    during tunneling in order to bypass the classical barrier, which
    physically corresponds to the increase of the variance of a state as it
    splits up into reflected and tunneled wave packets. The color scale is
    logarithmic with greens for negative values of the potential and blues for positive
    values. \label{f:Log}}
\end{center}
\end{figure}

We will use this all-orders closure in our analysis, illustrated in
Fig.~\ref{f:Log}, but will also see that it is beneficial to include an
additional quartic term in $s$ to bring the fourth-order moment,
$\Delta(\alpha^4)$, closer to its Gaussian value,
$\Delta(\alpha^4)=3s^4$ rather than $s^4$. (Similar parameterizations of moments have been used
in other cosmological analyses, such as \cite{QuantumHiggsInflation,EffPotInflation}.) The
constraint then reads
\begin{equation} \label{UAllFour}
   0=p_{\alpha}^2+p_s^2+\frac{U}{s^2}
   +\frac{1}{2}\left(U_p(\alpha+s)+U_p(\alpha-s)\right)+ \frac{1}{12} U_p^{''''}(\alpha)s^4\,.
\end{equation}
(The last term equals $2U_p''''s^4/4!$, which increases the fourth-order term
$U_p''''s^4/4!$ contained in the all-orders contribution to the Gaussian value
of $3U_p''''s^4/4!$.)  Additional amendments at higher moment orders may also
be considered, but they will not play a large role in the first analysis
presented here.

\subsection{Approximations}

A characteristic qualitative feature of the extended potential
$\frac{1}{2}(U_p(\alpha+s)+U_p(\alpha-s))$ is an extension of the classical
confined region to a channel that reaches smaller $\alpha$ for larger $s$; see
Fig.~\ref{f:Log}: If the local minimum of the potential in the confined region
is located at $\alpha_0$, the extended potential at $(\alpha,s(\alpha))$ along the
line $s(\alpha)\approx \alpha_0-\alpha$ is much smaller than the classical
potential $U_p(\alpha)$ because $U_p(\alpha+s(\alpha)\approx U_p(\alpha_0)$ is
much smaller than $U_p(\alpha)$ as well as $U_p(\alpha-s)$. The classical
confined region is therefore extended into a channel along
$s(\alpha)\approx \alpha_0-\alpha$ in the $(\alpha,s)$-plane. The additional
quartic contribution in (\ref{UAllFour}) preserves the channel and only
modifies its width for small $s$. (See Fig.~\ref{f:ChannelFour} below.)

Properties of the channel are important for tunneling
dynamics. Characteristic features are given by the zero levels of the extended
potential in the $(\alpha,s)$-plane as well as the $s$-dependent location of
the local maximum in the $\alpha$-direction. The potential and its
$\alpha$-derivative are polynomials in $\exp(\alpha)$ of higher than quadratic
order, such that exact expressions for the zero levels and local maxima would
be hard to find, or lengthy. Fortunately, since the walls of the channel are
rather steep for common parameter choices, the subtraction of $p^2$ does not
significantly change the zero levels, and it does not change local extrema at
all. For sufficiently large $s$, we can ignore the $1/s^2$-term. Moreover,
around the barrier, whose properties are described by the local maxima as well
as one of the zero levels at smaller $\alpha$, the $\gamma$-term in the
potential can be ignored because $\alpha$ is negative in this region. The
right-most zero level is located at positive $\alpha$, where the $k$-term in
the potential can be ignored.

\begin{figure}
\begin{center}
  \includegraphics[width=13cm]{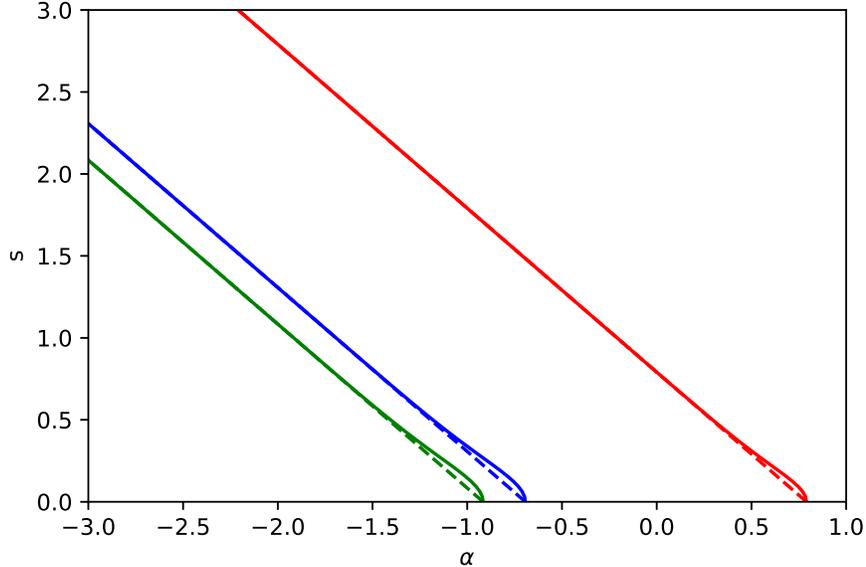}
  \caption{Characterization of the channel, given by the local maxima in the
    $\alpha$-direction (green), the left side of the channel (blue) and its
    right side (red). Dashed lines show the approximations
    (\ref{alphaexp}). The relevant parameters are $k=1$ and
    $\gamma=1.05$. \label{f:Contours}}
\end{center}
\end{figure}

These approximations lead to simple equations
for the desired quantities, given by
\begin{equation}
  e^{\alpha_{\rm max}(s)} = \frac{2k}{5} \frac{\cosh(4s)}{\cosh(5s)}
\end{equation}
for the local maxima in the $\alpha$-direction,
\begin{equation}
  e^{\alpha_{\rm left}(s)} = \frac{k}{2} \frac{\cosh(4s)}{\cosh(5s)}
\end{equation}
for the left zero level, and
\begin{equation}
  e^{\alpha_{\rm right}(s)} = 2\gamma^2\frac{\cosh(4s)}{\cosh(5s)}
\end{equation}
for the right zero level. For relatively large $s$, these equations can be
simplified further by using $\cosh(x)\approx \frac{1}{2}e^x$ for $x\gg
1$. Thus, we arrive at
\begin{equation} \label{alphaexp}
  \alpha_{\rm max}(s)\approx \ln(2k/5)-s\quad,\quad \alpha_{\rm left}(s)
  \approx \ln(k/2)-s\quad,\quad \alpha_{\rm right}(s)\approx
  \ln(2\gamma^2)-s\,.
\end{equation}
Figure~\ref{f:Contours} demonstrates the reliability of these approximations.

\begin{figure}
\begin{center}
  \includegraphics[width=14cm]{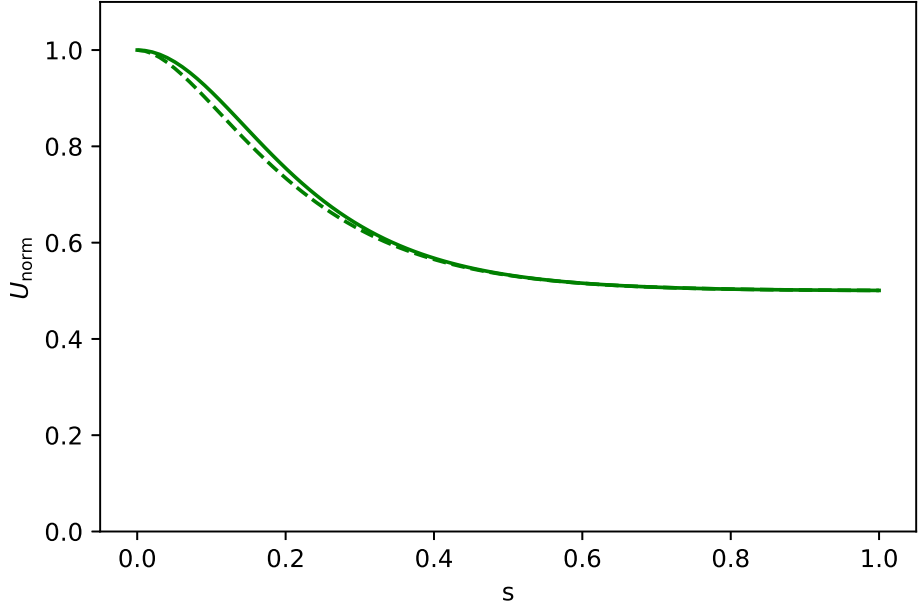}
  \caption{Behavior of the barrier height as a function of $s$, normalized to
    unit height of the classical barrier: $U_{\rm norm}(s)=(U_p(\alpha_{\rm
      max}(s),s)+p^2)/(U_{\rm max}+p^2)$. The dashed line is based on the
    simplified expression of $\alpha_{\rm max}(s)$ in (\ref{alphaexp}). \label{f:Max}}
\end{center}
\end{figure}

The classical local maximum, located at $\alpha_{\rm max}(0)=\ln(2k/5)$, has a
height of
\begin{equation}
  U_{\rm max}=\frac{k}{5\beta^2} \left(\frac{2k}{5}\right)^4-p^2\,,
\end{equation}
using the same approximation as in the derivation of $\alpha_{\rm max}(s)$.
The value of the extended potential decreases along $\alpha_{\rm max}(s)$. In
the limit of very large $s$, we can ignore contributions from $\exp(\alpha-s)$
in the extended potential while $\exp(\alpha+s)$, according to
(\ref{alphaexp}) becomes independent of $s$. With these approximations, we can
see that the maxima in the $\alpha$-direction approach a constant value which
turns out to equal
\begin{equation}
   \lim_{s\to\infty} \left(U_p(\alpha_{\rm max}(s),s)+p^2\right)= \frac{k}{10\beta^2}
   \left(\frac{2k}{5}\right)^4= \frac{1}{2} \left(U_{\rm max}+p^2\right)\,.
\end{equation}
For the barrier to disappear by quantum effects for the class of states
described by our closure condition, we therefore need
$p^2>\frac{1}{2} (U_{\rm max}+p^2)$, or $p^2>U_{\rm max}$. The full dependence
of the local maxima in the $\alpha$-direction is shown in Fig.~\ref{f:Max}.
 
\subsection{Features of tunneling trajectories}

We have numerically analyzed evolution in our versions of quantum potentials,
using rather small values for $\gamma$ in order to avoid steep potential walls
on which reflections of an evolving trajectory are hard to resolve. The case
of cosmological interest would rather be large values of $\gamma$ that imply a
large confined region which models long-term expansion of a universe. The
small values of $\gamma$ used here nevertheless allow us to infer interesting
qualitative features of trajectories that are expected to hold also for large
$\gamma$.

\begin{figure}
\begin{center}
  \includegraphics[width=12cm]{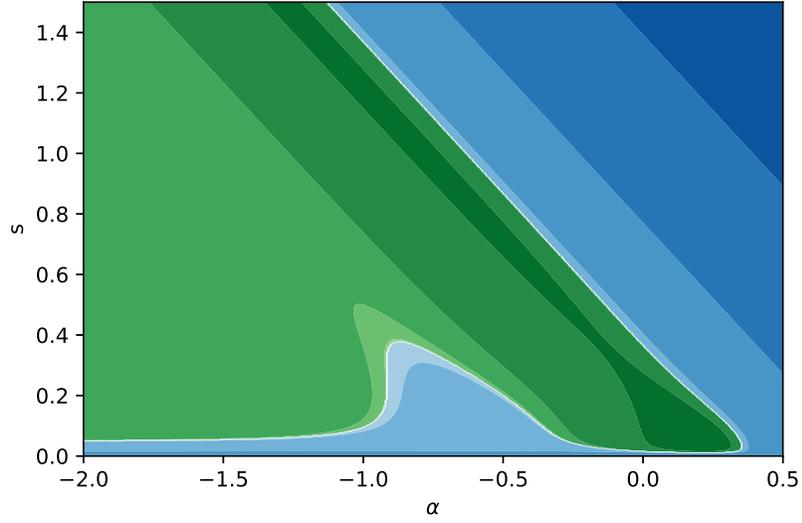}
  \caption{Larger range of the quantum potential with the same parameters and
    color scale as in Fig.~\ref{f:Log}. \label{f:Channel}}
\end{center}
\end{figure}

Quantum potentials such as (\ref{USecond}), (\ref{UAll}) or (\ref{UAllFour})
show how tunneling dynamics can be realized in classical-type motion without
violating energy conservation. For instance, the second derivative in
(\ref{USecond}) is negative around a local maximum, and therefore the quantum
potential is lower than the classical barrier for non-zero $s$. In the present
case, the averaging of $U_p$ at $\alpha+s$ and $\alpha-s$ contained in
(\ref{UAll}) and (\ref{UAllFour}) not only implies a similar lowering of the
barrier, as shown in Fig.~\ref{f:Channel}, but also extends the classically
oscillating region into a channel that reaches to negative values of $\alpha$
for sufficiently large $s$; see Fig.~\ref{f:Log}.

\begin{figure}
\begin{center}
  \includegraphics[width=15cm]{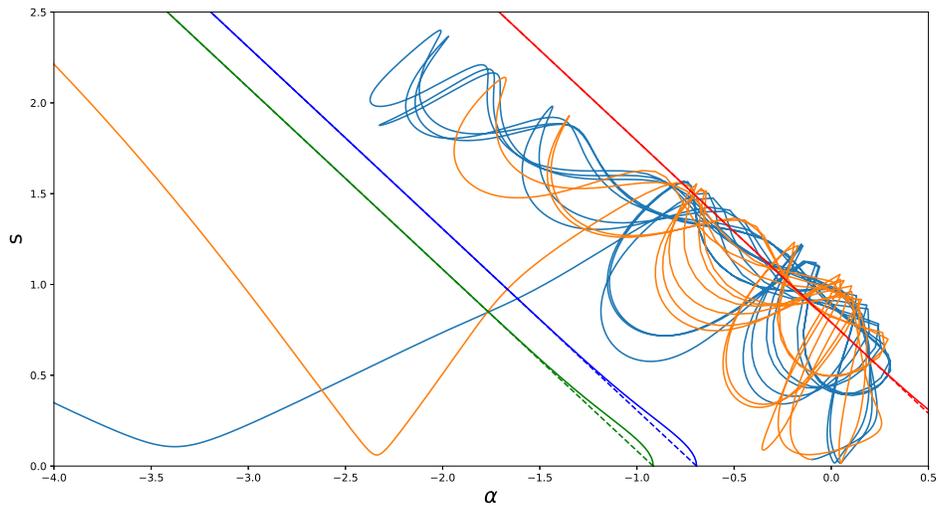}
  \caption{Tunneling trajectory in the amended all-orders potential
    (\ref{UAllFour}). The diagonal lines indicate characteristic features of
    the non-amended wall as in Fig.~\ref{f:Contours}. The additional quartic
    contribution in the amended potential allows the trajectory to penetrate
    the right wall of the channel at small $s$. Blue and orange parts of the
    trajectory indicate times before and after a random initial condition, respectively.
    \label{f:ChannelTrajectory}}
\end{center}
\end{figure}

A negative potential at zero energy does not necessarily imply that a
trajectory can cross the barrier if there are more than one dimension. In one
dimension, the momentum is non-zero under these conditions and the object
keeps moving in the same direction, but in two or more dimensions an object
can get deflected and turn around while its momentum remains non-zero. Based
on numerical simulations with random initial values that start around
$\alpha=0$ and small $s$, we have found that trajectories often get stuck in
the channel and keep moving along it to larger and larger $s$. At such large
$s$, the channel is very straight such that a trajectory, once it reaches this
region, follows a periodic pattern between deflections at the channel walls
without moving out. While the channel guides the trajectory toward smaller
$\alpha$, very far to the left of the classical barrier, we do not consider
these solutions to be good examples of tunneling because our quasiclassical
approximation and the moment closure become unreliable at large $s$.

\begin{figure}
\begin{center}
  \includegraphics[width=11cm]{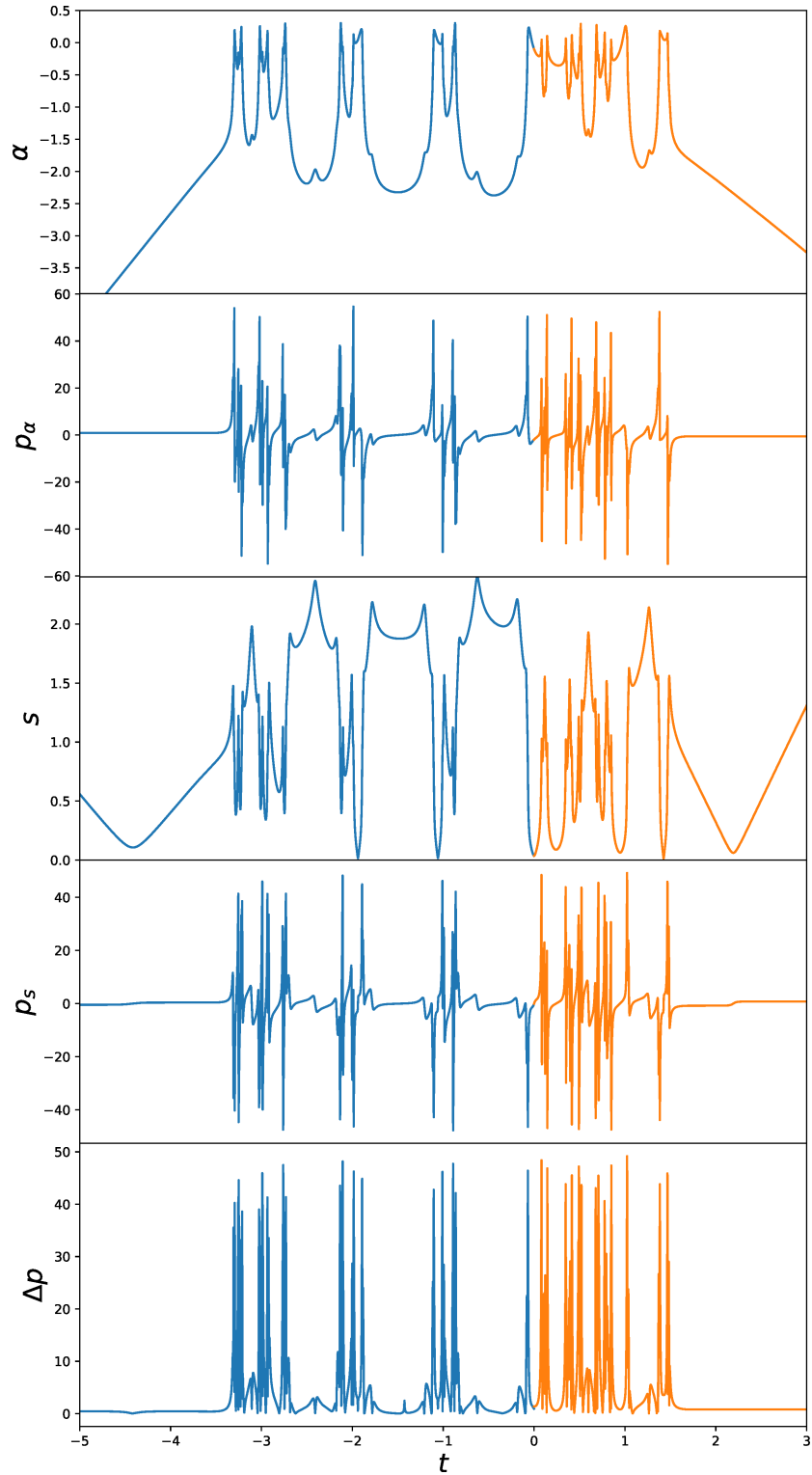}
  \caption{Details of the tunneling trajectory shown in
    Fig.~\ref{f:ChannelTrajectory}. A random initial condition has been set
    where the blue and orange curves meet.
\label{f:Trajectory}}
\end{center}
\end{figure}

\begin{figure}
\begin{center}
  \includegraphics[width=12cm]{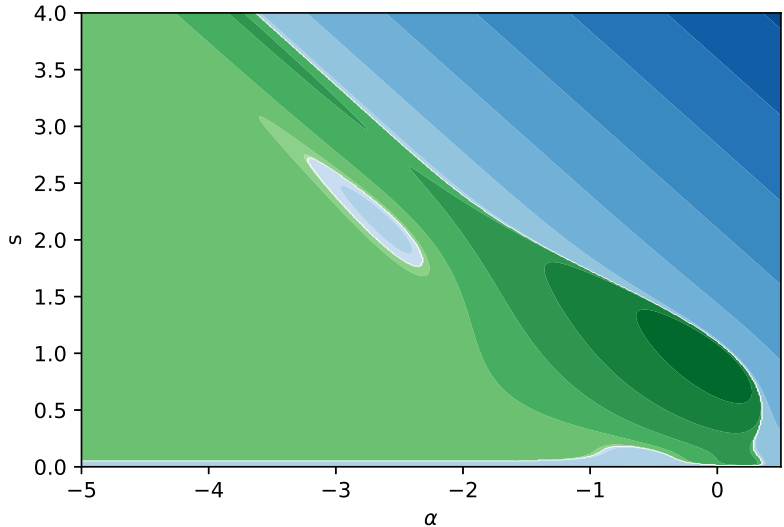}
  \caption{The amended potential (\ref{UAllFour}) with the same parameters as
    in Fig.~\ref{f:Log}. \label{f:ChannelFour}}
\end{center}
\end{figure}

\begin{figure}
\begin{center}
  \includegraphics[width=12cm]{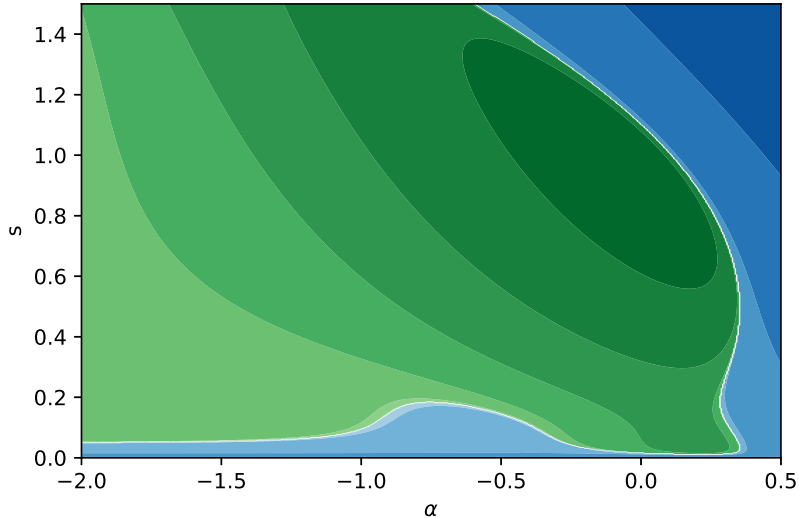}
  \caption{Closer view of the end of the channel from
    Fig.~\ref{f:ChannelFour}. \label{f:ChannelFourSmall}}
\end{center}
\end{figure}

Deviations from the all-orders closure, as implied by contributions from
higher moments different from (\ref{Closure}), make the channel irregular,
such that a trajectory bounces off the channel walls at different angles each
time, making it more likely to exit the channel eventually. At very large $s$,
several higher-order moments are relevant in an amended closure. It is then
hard to derive generic information without prior knowledge of the tunneling
state and its moments. Fortunately, as shown by
Figs.~\ref{f:ChannelTrajectory} and \ref{f:Trajectory}, the fourth-order
amendment of the potential in (\ref{UAllFour}) makes it possible to find
trajectories that enter and exit the channel at rather small $s$. The
additional fourth-order contribution to the potential, shown in
Figs.~\ref{f:ChannelFour} and \ref{f:ChannelFourSmall}, keeps the trajectory
closer to the end of the channel at smaller $s$, where it has several
opportunities to probe the channel wall under different impact directions and
eventually crosses the barrier.

As can be seen in Figure~\ref{f:ChannelTrajectory}, $s$ reaches large values
also to the left of the classical barrier, where this variable keeps
increasing after a single reflection at small $s$, caused by the $U/s^2$
contribution to the potential. This increase to large values is expected
because the classical potential is nearly constant in this region. The
trajectory therefore behaves like the quantum fluctuation $s=\Delta\alpha$ of
a free particle, which increases before and after its minimum value,
increasing linearly for asymptotically large times.

It is noteworthy that the minima of $s$ in the nearly free region are located
close to the classical barrier before and after tunneling in and out of the
trapped region. This behavior is expected if one imagines a wave function
approaching a barrier, such that it gets more narrow as some of its front
part starts getting reflected back toward the center. Similarly, a wave packet
that tunnels out of the trapped region may narrow down briefly when only a small
contribution is left in the trapped region. Based on the extended potential as
a function of $\alpha$ and $s$, the minima of the free region are generically
located near the barrier because the trajectory has to approach the channel
wall, located between the blue and green lines in the figure, at close to a right
angle. Under this condition it is able  to move through and escape, rather
than being deflected back into the channel. The direction of the channel
implies that escaping trajectories are aimed toward smaller $s$, toward the
$U/s^2$-potential where they reach their local minima.

\begin{figure}
\begin{center}
  \includegraphics[width=16cm]{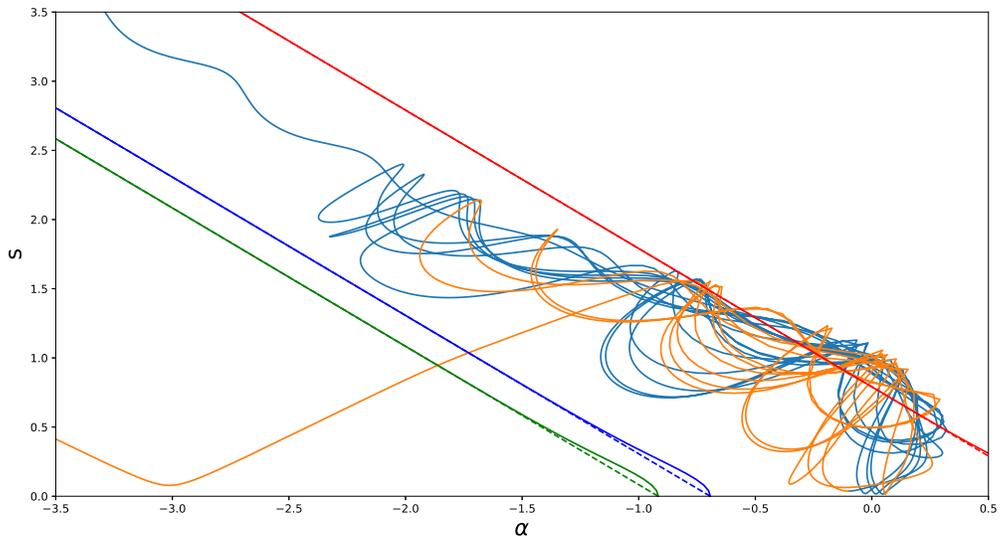}
  \caption{Tunneling trajectory in the amended all-orders potential
    (\ref{UAllFour}). All parameters are the same as in
    Fig.~\ref{f:ChannelTrajectory}, except that the initial value of $\alpha$
    has been changed from $-0.09938693911314142$ to $-0.09938693911314141$, a
    difference only in the last relevant decimal place. The final outcome of
    this tiny change is very large because the trajectory now gets stuck in
    the channel. 
\label{f:ChannelTrajectory2}}
\end{center}
\end{figure}

Our quasiclassical trajectories therefore provide a meaningful and geometrical
description of the beginning and the end of a tunneling process. The trapped
part of the trajectory is harder to interpret, but it is clear that it is much
more complicated than the classical solution in this region, which at constant
energy would oscillate with a regular period and amplitude. The combined
evolution of $\alpha$ and $s$, by contrast, has neither a regular period nor a
fixed amplitude, even though the quantum energy, given by our effective
Hamiltonian, is conserved. The complicated nature of quantum dynamics in the
trapped region is also highlighted by a high sensitivity to initial values, as seen
by comparing Fig.~\ref{f:ChannelTrajectory} with
Fig.~\ref{f:ChannelTrajectory2}.

The sensitivity to initial values is reminiscent of chaos, although we have
not performed a detailed analysis to demonstrate this feature. The classical
system is clearly non-chaotic (being 1-dimensional), but quantum dynamics may
nevertheless develop chaotic features as known for instance from Bohmian
treatments \cite{BohmianChaos}. Another indication that the extended dynamics
here may be chaotic can be seen in the shape of the trapped region in the
amended all-orders potential, shown in Figs.~\ref{f:ChannelFour} and
\ref{f:ChannelFourSmall}. As shown by the contours, the trapped region is
confined by walls that are partially concave, which may support chaotic
billiard motion as in other cosmological models, such as anisotropic ones
\cite{Billiards}.

\section{Conclusions}

Our analysis of extended quantum potentials has suggested a strategy to find
and study quasiclassical tunneling solutions for an oscillating universe
model. We derived quantum corrections to the classical potential based on an
assumption about the moment closure of states. A closure condition is unlikely
to describe all relevant states, but it can reveal some properties of
dynamical tunneling, provided solutions stay in regions in which the closure
condition can be considered a good approximation. For instance, a precise
closure at higher orders of moments should not be required if the fluctuation
variable, $s$, remains sufficiently small.

While a full quantum treatment would imply that any state initially supported
in the oscillating region will eventually tunnel and approach the singularity
at $\alpha\to-\infty$, perhaps after separating into several wave packets that
had tunneled at different times, our quasiclassical description implies
tunneling only under certain conditions on the initial values of a trajectory.
In particular, although the value of $s$ should not become too big for our
approximations to be valid, it has to grow sufficiently large close to the
classical barrier for the local maximum in the $\alpha$-direction (one of the
channel walls) to have dropped below zero. This condition cannot be fulfilled
for all parameter values but requires, in partiular, that the parameter $p$
that determines the asymptotic potential at $\alpha\to-\infty$ is sufficiently
large. If $p$ is too small, we would not see any quasiclassical tunneling
solutions in our model even though quantum tunneliung in a full treatment
would certainly occur. Quasiclassical models of the form considered here,
therefore, cannot provide a complete description of tunneling. But solutions
that stay within the allowed ranges of parameters may still provide
interesting dynamical information that would be harder to find using
traditional methods.

Our main result is the observation that oscillations in the trapped region,
seen over many cycles, are much less regular in the quantum case than they
appear classically. Our numerical simulations were restricted to small
$\gamma$, or rather narrow trapped regions, because the steep potential walls
implied by larger values of this parameter make it hard to achieve reliable
numerics. Qualitatively, larger $\gamma$ imply longer cycles in the trapped
region, which may give the appearance of more regular behavior because the new
quantum variable, $s$, does not change as abruptly during a single cycle as it
does after multiple reflections off the potential walls. Nevertheless, the
succession of several cycles should also be less regular than in the classical
case if $\gamma$ is large because each cycle generically starts with different
values of $s$ and $p_s$, which affect the evolution of $\alpha$ through
quantum back-reaction. A single model may therefore probe a large number of
cosmological cycles with different maximal expansion, even if matter
parameters remain the same.

A quasiclassical model may also be crucial in developing a scenario that couples
the isotropic background to perturbative anisotropies or inhomogeneity. Such a
combination would be harder to analyze at the full quantum level where a
combined wave function for background and inhomogeneity would have to be
evaluated. It would be easier, by comparison, to couple a quasiclassical
background model to a standard description of perturbative inhomogeneity and
analyze how quantum effects could affect the evolution of inhomogeneous
modes through a tunneling process. 

\section*{Acknowledgements}

This work was supported in part by NSF grant PHY-1912168.


\end{document}